\documentclass[floatfix,aps,nofootinbib,showpacs]{revtex4}

\usepackage[xdvi]{graphicx}
\usepackage{amsmath}
\usepackage{enumerate}

\begin{document}
\title{The Hubble Slow Roll Expansion for  Multi Field Inflation}
\author{Richard Easther}\email{richard.easther@yale.edu} \affiliation{Department of Physics, Yale University, New Haven  CT 06520, USA}  

\author{John T. Giblin, Jr}\email{jtg@het.brown.edu}\affiliation{Department of Physics, Brown University, Providence  RI  02912, USA}

\begin{abstract}
We examine the dynamics of inflation driven by multiple, interacting scalar fields and derive a multi field version of the Hubble slow roll expansion.    We show that the properties of this expansion naturally generalize those of the single field case. We  present an analogous hierarchy of slow roll parameters, and derive the system of ``flow'' equations that describes their evolution, and show that when this system is truncated at finite order, it can always be solved exactly.    Lastly, we express the scalar and tensor perturbation spectra in terms of the slow roll parameters. 
\end{abstract}

\maketitle 

\section{Introduction}

Scalar field driven inflation \cite{guth} solves the traditional problems (horizon, flatness, structure) of early universe cosmology. Fortuitously,  the need for an almost scale invariant perturbation spectrum strongly suggests that inflation is well described by the {\em slow roll\/} approximation \cite{steinhardtturner,linde82}: the parameters that describe the departures from scale invariance are also those that must be small if the slow roll approximation is to hold  \cite{liddleparsonsbarrow}.  In practice, we normally consider only the two lowest order slow roll parameters. However, from a theoretical perspective we recognize this is a truncated expansion and we are implicitly assuming that there are no large high-order derivatives in the dynamical system describing inflation.   

For a single field, the general slow roll expansion was written down by Liddle, Parsons and Barrow \cite{liddleparsonsbarrow}.\footnote{An analagous representation outside the context of inflation was introduced by Starobinsky \cite{starob1978}.}  The energy density and its proxy the Hubble parameter, $H$, is a function of the slowly rolling field, which serves as the independent variable of the dynamical system.    The slow roll expansion is expressed as a hierarchy of differential equations, involving progressively higher order derivatives of $H(\phi)$.  In many practical applications, this hierarchy is truncated at some finite order, and the properties of the system have been carefully studied in recent years.  In particular, Kinney \cite{kinney02} showed that this system possesses a set of attractors.  Easther and Kinney \cite{eastherkinney}  built on this observation, using the slow roll hierarchy to generate random models of inflation, and this technique was later employed in the analysis of the first year of WMAP data \cite{peiris,wmap} and more recently in \cite{fengbo,Efstathiou:2005tq}. However, the randomly generated models are not associated with a measure so one cannot quantify which region of parameter space is favored.  Subsequently,  Liddle pointed out that the truncated slow roll hierarchy can always be solved analytically \cite{liddle03}.  Leach and Liddle recently used this solution to implement and test different reconstruction algorithms, and their results illustrate how the lack of a measure makes it very difficult to draw quantitative conclusions about the parameter space of slow roll inflation \cite{ramirezliddle}.

The developments above all refer to models of inflation driven by a single scalar field.   However, multi field models are common in the literature and have a number of attractive features. Firstly, there is no theoretical guarantee that any inflationary phase can be described in terms of a single evolving field: the primarily motivation for doing so is simply an application of Occam's Razor.  Moreover, many phenomenologically promising models of inflation (e.g. \cite{linde94,copelandliddle}) require two or more interacting fields. Lastly, recent developments in string theory have motivated the study of multi field models in the context of the string landscape \cite{kklmmt,easther04,burgess,barnard}.

The purpose of this paper is to generalize the Hubble slow roll formalism to models with multiple fields.  Section \ref{singlefield} recaps  the slow roll regime for the single field case and Section \ref{multifield} gives the multi field version, which is the main result of this paper.   We see that almost all features of the single field case carry over to the multi field case.   In particular, the exact solution found by Liddle for the truncated single field slow roll hierarchy generalizes to the multi field case, and we write  down an exact algebraic expression for the potential which corresponds to the truncated slow roll hierarchy at arbitrary order.   Finally we express the scalar and tensor perturbation spectra in terms of the generalized slow roll parameters. 

\section{Single Field Formalism}
\label{singlefield}
\subsection{Background Evolution}

In scalar field driven inflationary models, the inflaton field is governed by the following equation of motion,
\begin{equation}\label{1deom}\ddot{\phi}+3H \dot{\phi}+\frac{d V}{ d \phi}=0.\end{equation}
The spacetime background obeys the Friedman equations
\begin{equation}
\label{fried1}
H^2=\frac{8 \pi G}{3}\left[{V(\phi)+\frac{1}{2}\dot{\phi}^2}\right]
\end{equation}
and
\begin{equation}
\label{fried2}
\frac{\dot{a}}{a}=\frac{8 \pi G}{3}\left[{V(\phi)-\dot{\phi}^2}\right],
\end{equation}
where $H=\frac{\dot{a}}{a}$ is the  Hubble parameter. Equations  (\ref{fried1}) and (\ref{fried2}) can be combined to produce the Hamilton-Jacobi equation \cite{salopekbond}
\begin{equation}
\label{H2epsilon}
H^2(\phi)\left[{1-\frac{1}{3}\epsilon(\phi)}\right]=\frac{8 \pi G}{3}V(\phi),
\end{equation}
where the {\it slow roll parameter} $\epsilon$ is   \cite{copelandkolb,liddleparsonsbarrow}
\begin{equation}
\label{epsilon}
\epsilon = \frac{1}{4 \pi G} \left(\frac{1}{H}\frac{d H}{d\phi}\right)^2.
\end{equation}
Note one frequently sees the  {\it potential slow roll} expansion which can be obtained directly from the functional form of the potential \cite{liddleparsonsbarrow,liddlelyth}
\begin{eqnarray}
\epsilon_V &=& \frac{1}{16\pi G}\left({\frac{V^\prime(\phi)}{V(\phi)}}\right)^2 \\
\eta_V &=& \frac{1}{8\pi G} \frac{V^{\prime \prime}(\phi)}{V(\phi)}.
\end{eqnarray}

We can understand the physical significance of $\epsilon$ by rewriting (\ref{fried2}) as
\begin{equation}
\label{fried3}
\left({\frac{\ddot{a}}{a}}\right)=H^2\left[{1-\epsilon}\right],
\end{equation}
and we immediately see  that $\epsilon<1$  corresponds to $\ddot{a}>0$, the necessary condition for inflation.

\subsection{Hierarchy of Parameters}

The duration of inflation is most frequently parameterized by $N$, the number of e-foldings by which the scale factor grows during the inflationary phase, so $a \equiv e^N$. As a matter of definition, we obtain $N$ by integrating $H$,
\begin{equation}
\label{1dN}
N=\int_t^{t_e} H dt.
\end{equation}
As written above, $N$ measures the number of e-folds {\em before\/} the end of inflation, and increases as $t$ decreases.  We can write $N$ as a function of $\phi$ via 
\begin{equation}
\label{1dN2}
N=  \int_\phi^{\phi_e} \frac{H}{\dot{\phi}} d\phi= {2\sqrt{\pi G}} \int_{\phi_e}^\phi \frac{d\phi}{\sqrt{\epsilon(\phi)}}.
\end{equation}
We now write a differential equation for $\epsilon (N)$,  
\begin{equation}
\frac{d\epsilon}{dN} = \epsilon(\sigma+2\epsilon),
\end{equation}
where 
\begin{equation}
\label{1flow1}
\sigma =\frac{1}{\pi\sqrt{G}}\left[{\frac{1}{2}\left({\frac{\partial^2_\phi H}{H}}\right)-\left(\frac{\partial_\phi H}{H}\right)^2}\right].
\end{equation}
Equation (\ref{1flow1}) is the first {\it flow equation} \cite{kinney02} and $\sigma$ depends on the second  derivative of $H$.  Defining additional parameters \cite{liddleparsonsbarrow} 
\begin{eqnarray}
\label{1dparameters}
^n\lambda &=& \left({\frac{1}{4 \pi G}}\right)^n\frac{(\partial_\phi H)^{n-1}}{H^n}\partial_\phi^{n+1}H,
\end{eqnarray}
we create an infinite system of coupled first order differential equations:\footnote{Note that $\sigma=2(\,^1\lambda)-4\epsilon$}
\begin{eqnarray}
\frac{d\sigma}{dN} &=& -\epsilon(5\sigma+12 \epsilon)+2(\,^2\lambda)\\
\frac{d\,^n\lambda}{dN} &=& \,^n\lambda\left[{\frac{1}{2}(n-1)\sigma+(n-2)\epsilon}\right]+\,^{n+1}\lambda .
\end{eqnarray}

Looking at  (\ref{1dparameters}) we see that the derivative of  $\,^m\lambda$ depends on four (or fewer) quantities: $\,^m\lambda$, $\,^{m+1}\lambda$, $\sigma$, and $\epsilon$.  These flow equations can be truncated to a system of finite order if for some $n$, $^n\lambda=0$, for all $N$.  Once we have specified the values of the $\,^n\lambda$ at some point $N$ we can then calculated the full inflationary evolution as a function of $N$ and then write down an expression for  $V(\phi)$.

\section{Multiple Fields}
\label{multifield}
\subsection{Background Evolution}

We now want to extend the Hubble slow roll expansion  to an inflationary model driven by $M$ scalar fields, $\phi_i$.  The Friedman equations (\ref{fried1},\ref{fried2}) become
\begin{eqnarray}
\label{mfried1}
H^2&=&\frac{8 \pi G}{3}\left[{V({\phi})+\sum_{i=0}^M\frac{1}{2}\dot{\phi_i}^2}\right] \\
\label{mfried2}
\frac{\ddot{a}}{a}&=&\frac{8 \pi G}{3}\left[{V({\phi})-\sum_{i=0}^M\dot{\phi_i}^2}\right]  \, .
\end{eqnarray}
The single equation of motion (\ref{1deom}) turns into $M$ equations
\begin{equation}
\label{multeom}
\ddot{\phi_i}+3H \dot{\phi_i}+\frac{\partial V(\phi)}{ \partial \phi_i}=0 \, ,
\end{equation}
where an $\phi$ is now being used as a shorthand for all $M$ $\phi_i$.  To produce multi field analogs of (\ref{epsilon}) and (\ref{fried3}) we follow the method of \cite{copelandkolb}.  Firstly, the time derivative of (\ref{mfried1}) is
\begin{equation}
2H \frac{dH}{dt}=\frac{8 \pi G}{3} \sum_{i=1}^M \frac{d\phi_{i}}{dt}\left({\frac{d^2\phi_i}{dt^2} +\frac{\partial V(\phi)}{\partial \phi_i} }\right),
\end{equation}
which, when combined with the corresponding equation of motion (\ref{multeom}), produces
\begin{equation}
2 \frac{dH}{dt}=-{8 \pi G}\sum_{i=1}^M \left(\frac{d\phi_i}{dt}\right)^2. 
\end{equation}
In general a time derivative can be changed into a derivative with respect to the fields by 
\begin{equation}
\frac{d}{dt}=\sum_{j=1}^M \frac{d\phi_j}{dt}\frac{\partial}{\partial \phi_j} ,
\end{equation}
and so
\begin{equation}
2\sum_{j=1}^M \frac{\partial H}{\partial \phi_j} \frac{d\phi_j}{dt}=-{8 \pi G}\sum_{i=1}^M {\left(\frac{d\phi_i}{dt}\right)^2}.
\end{equation}
Since the fields are independent, corresponding terms can be set equal, 
\begin{equation}
\label{evolution}
\dot{\phi_j}=-\frac{1}{4 \pi G} \frac{\partial H}{\partial \phi_j}.
\end{equation}
This  defines the unique relationship between the temporal derivative of a given field and the derivative of the Hubble parameter with respect to that field.

We use this result to rewrite the equation of motion as \cite{salopekbond, copelandkolb}  
\begin{equation}
 H^2 = \frac{1}{12 \pi G} \sum_{i=1}^M \left({\frac{\partial H}{\partial \phi_i}}\right)^2 + \frac{8 \pi G}{3} V 
\end{equation}
or
\begin{equation}
\label{multihj}
H^2\left[{1-\frac{1}{H^2}\frac{1}{12 \pi G} \sum_{i=1}^M \left({\frac{\partial H}{\partial \phi_i}}\right)^2}\right]=\frac{8 \pi G}{3} V.
\end{equation}

We define a new set of parameters $\epsilon_i$, such that 
\begin{equation}
\label{epsilontilde}
\epsilon_i = \sqrt{\frac{1}{4\pi G}}\frac{1}{H}\frac{\partial H}{\partial \phi_i}.
\end{equation}
In practice we will frequently look at the sum of squares, which can be written as a dot product:
\begin{equation}
\label{mepsilon}
\epsilon=\sum_{i=1}^M\epsilon_i \epsilon_i=\sum_{i=1}^M{\frac{1}{4\pi G}}\frac{1}{H^2}\left({\frac{\partial H}{\partial \phi_i}}\right)^2 \, .
\end{equation}
In principle, we should denote this quantity $\epsilon^2$ but in practice we use $\epsilon$ to emphasize the connection with the single field slow roll expansion, which is recovered by setting $M=1$. Substituting (\ref{mepsilon}) into (\ref{multihj}) we see
\begin{equation}
\label{multifriedmann}
H^2(\phi)\left[{1-\frac{1}{3} }\epsilon  \right]=\frac{8 \pi G}{3}V(\phi),
\end{equation}
which is identical in form to the single field analog (\ref{H2epsilon}), although $\phi$ and $\epsilon$ are now understood to have their multi-component forms.   Similiarly, we can rewrite (\ref{mfried2}) as 
\begin{equation}
\label{addot}
\left(\frac{\ddot{a}}{a}\right) = H^2\left[{1-\epsilon}\right],
\end{equation}
confirming that $\epsilon<1 \iff \ddot{a}>0$ for the multi field case.

\subsection{The Multi Field Slow Roll Hierarchy} 
\label{parameters}

Now that we have a well-defined parameter $\epsilon$, we can   build our slow roll hierarchy. We start with
\begin{equation}
\label{epep}
\frac{d{\epsilon_i}}{dN}= \sqrt{\frac{1}{4 \pi G}} \left[{ \frac{d}{dN}\left(\frac{\partial  H}{\partial \phi_i}\right) \frac{1}{H} - \frac{dH}{dN} \frac{\partial H}{\partial \phi_i}\frac{1}{H^2} }\right]
\end{equation}
or in terms of $\frac{d}{dN}=\sum\frac{d\phi_i}{dN}\frac{\partial}{\partial \phi_i}$ and $\frac{dH}{dN}=\epsilon H$,
\begin{equation}
\frac{d{\epsilon_i}}{dN}= \sqrt{\frac{1}{4 \pi G}} \left[{ \sum_{j=1}^M\frac{d\phi_j}{dN}\frac{\partial}{\partial \phi_j} \left(\frac{\partial  H}{\partial \phi_i}\right) \frac{1}{H} - \epsilon \frac{\partial H}{\partial \phi_i}\frac{1}{H} }\right].
\end{equation}
We define our first evolution equation:
\begin{equation}
\label{flow1}
\frac{d \epsilon_i}{d N} = \left[{\sum_{j=1}^M\epsilon_j\,^1\lambda_{ij} - \epsilon_i\epsilon }\right],
\end{equation}
where $^1\lambda_{ij}$ a higher order parameter defined by 
\begin{equation}
\label{1lambda}
^1\lambda_{ij} = \left({\frac{1}{4 \pi G}}\right) \frac{1}{H}\frac{\partial^2 H}{\partial \phi_i \partial \phi_j}.
\end{equation}
Additional parameters are a generalization of (\ref{1lambda}):
\begin{equation}
\label{mlambda}
^m\lambda_{\alpha_0 \ldots \alpha_m} = \left(\frac{1}{4\pi G}\right)^m \frac{1}{H^m} \left[{\prod_{i=2}^m \left(\frac{\partial H}{\partial \phi_{\alpha_i}}\right)}\right] \frac{\partial^{m+1} H}{\partial \phi_{\alpha_0} \ldots \partial \phi_{\alpha_m} }.
\end{equation}
Note that the product in ({\ref{mlambda}) begins at $2$ and not $0$; the first two indices are not  included in this product.  

Using these higher order parameters, we can then calculate a system of coupled first order differential equations,
\begin{align}
\label{flow2}
\frac{d \,^m\lambda_{\alpha_0 \ldots \alpha_m}}{dN} = & \sum_{i=1}^M \,^{m+1}\lambda_{\alpha_0\ldots\alpha_m i} \, + ^{m}\lambda_{\alpha_0\ldots\alpha_m}\left({\sum_{i=1}^M \sum_{k=2}^m \frac{^1\lambda_{i\alpha_k}}{\epsilon_{\alpha_k}} \epsilon_i -m \epsilon }\right)
\end{align}
where we've used the relationship $d\phi_i / dN$ by looking at
\begin{equation}
\label{N}
N=\int_t^{t_e} H dt = \int_{\phi_i}^{\phi_{i}(e)} \frac{H}{\dot{\phi}_i} d\phi_i= {\sqrt{4 \pi G}} \int_{\phi_i(e)}^{\phi_i} \frac{d\phi_i}{{\epsilon_i}},
\end{equation}
where the third equality is justified by the definition of $\epsilon_i$ (\ref{epsilontilde}) and (\ref{evolution}).


We now explore the properties of this hierarchy.  First, when $M=1$ we have a single field and the lower indices become redundant; we drop them by 
\begin{equation}
^m\lambda_{i \ldots i} = {}^m\lambda.
\end{equation}
It is then trivial to check that we then recover the single field flow equations.

These parameters  exhibit several useful symmetries.  Looking at (\ref{mlambda}) we see that the $\lambda$  are symmetric in the {\it first two} indices and separately symmetric in every other index.  We can further reduce the number of independent parameters by recognizing that interchange of one the first two indices, $\alpha_i$, with one of the other indices, $\alpha_j$, is equivalent to multiplying by a factor of $\epsilon_{\alpha_i}/\epsilon_{\alpha_j}$.  This is easily verified by a straightforward calculation. Consequently, we find that any two $^m\lambda$ can be obtained in terms of one another (and the $\epsilon_i$) if they possess the same set of numerical subscripts as each other.  At root, this corresponds to an implicit assumption that $H$ is a well behaved function of the $\phi_i$ and that all mixed derivatives commute.   The final ambiguity that arises in the multi field case is that we are always free to rotate or translate the fields in any way, provided that the kinetic terms retain their canonical form.  This will necessarily redefine the $\lambda$'s, although the overall physics remains unchanged. This issue does not arise in the single field case, since there is only one field and it cannot be transformed in a nontrivial way without changing the form of the kinetic term.

\subsection{Alternative Parameters}
\label{alternateparameters}

In developing a previous formalism for perturbations in multiple field slow roll inflation, Nibbelink and van Trent introduce the following parameters \cite{nibbelink1, nibbelink2}:
\begin{eqnarray}
\tilde{\epsilon}(\phi) &=& - \frac{\dot{H}}{H^2} \\
\tilde{{\bf \eta}}^{(n)}(\phi) &=& \frac{\phi^{(n)}}{H^{n-1}\left|{\dot{\phi}}\right|},
\end{eqnarray}
where we use the ordinary time convention of \cite{nibbelink1} and the notation of \cite{nibbelink2}.  We can quickly show that the lowest order parameter used in their expansion is the same as our $\epsilon$:
\begin{equation}
\tilde{\epsilon} = - \frac{\dot{H}}{H^2} = - \sum_{i=1}^M \frac{1}{H} \frac{\partial N}{\partial \phi_i} \frac{\partial H}{\partial \phi_i} = \sum_{i=1}^M  \frac{\epsilon_i}{\sqrt{4 \pi G}} \frac{1}{H} \frac{\partial H}{\partial \phi_i} =\sum_{i=1}^M \epsilon_i^2 = \epsilon
\end{equation}
where our substitutions come from (\ref{N}).  The similarity here allows both treatments to reduce the Friedman equation to its convenient form (\ref{multifriedmann}). 

Our analyses diverge at this point.  Nibbelink and van Trent choose to differentiate with respect to a temporal parameter (either $t$ \cite{nibbelink1} or a more general temporal measure $\tau$ \cite{nibbelink2}), whereas with reconstruction in mind we choose to define the parameters with respect to derivatives of the field.    Since they both refer to the same physical system, the expressions obtained by Nibbelink and van Trent must be equivalent to ours. However, they differ in their relative tractability and in the clear connection between our formalism and the single field slow roll expansion.

\section{Evolution of Parameters and Reconstruction}

The infinite hierarchy of parameters introduced in Section \ref{parameters} provides a complete description of the evolution of $\epsilon = \epsilon_i \epsilon_i$.  We now follow the development of the single field analysis and truncate the hierarchy at finite order.  As before, we see that if all the $^m\lambda$ vanish for some $m$, then all the higher order parameters are also zero.  
To make this precise, assume there is some $\kappa$ such that for all $m > \kappa$  
\begin{equation}
\left.{\,^{m+1}\lambda_{\alpha_0 \ldots \alpha_{m+1}}}\right|_{\phi_i=0}=0.
\end{equation}
The derivatives of these parameters vanish, thanks to (\ref{flow2}).  What remains is a {\it closed} system of first order differential equations.

We can now reconstruct the inflationary potential, analogously to the procedure outlined in \cite{eastherkinney}. We begin by inverting (\ref{multifriedmann})
\begin{equation}
V=H^2\left({\frac{3}{8 \pi G}}\right)\left[{1-\frac{1}{3}\epsilon }\right].
\end{equation}
This requires knowledge of $H(N)$, $\epsilon(N)$, and $\phi(N)$.  We can trivially calculate $\epsilon$ from the flow equations, while to   track the evolution of the field one must look only at (\ref{evolution})
\begin{equation}
\label{dphidn}
\frac{d\phi_i}{dN}=-\sqrt{\frac{1}{4\pi G}}\epsilon_i.
\end{equation}
Using identities introduced previously, the differential equation for the Hubble parameter is  
\begin{equation}
\frac{dH}{dN}=\sum_{i=1}^M \frac{d\phi_i}{dN}\frac{\partial H}{\partial \phi_i}=H\sum_{i=1}^M \epsilon_i^2 =H\epsilon
\end{equation}
where our final unknown is the initial value of this parameter.  For this, we use the same normalization as in \cite{eastherkinney}; the amplitude of density fluctuations is on the order of $10^{-5}$ so
\begin{equation}
\label{hubblestart}
\frac{\delta \rho}{\rho} =\frac{H}{2 \pi}\sqrt{\frac{G}{\epsilon}} \, .
\end{equation}
From this it might appear that we have reconstructed the {\em full\/} potential. However, in  order to put this to use  we would need to measure a large number of $^m  \lambda$ with considerable precision, so   one cannot use this to reconstruct the inflationary dynamics in practice.

In \cite{kinney02} and
 \cite{eastherkinney} the single field flow equations were evolved numerically to recover the full inflationary dynamics after the slow-roll parameters were specified at a specific initial time.  Subsequently, Liddle pointed out that the flow equations can be solved analytically to produce $H(\phi)$ and $V(\phi)$.  This treatment yields the  inflationary potential in terms of the $^m\lambda$,  as evaluated at some initial time. We now show that a  similar truncation  works in the multi field case. Assume that $\,^{m+1}\lambda_{\alpha_0 \ldots \alpha_{m+1}}=0$ for all choices of $\alpha_i$.  Via  (\ref{mlambda}) and (\ref{flow2}), this corresponds to
\begin{equation}
 \frac{\partial^{m+1} H}{\partial \phi_{\alpha_0}\ldots \partial \phi_{\alpha_{m+1}} }=0 \, .
\end{equation}
The flow equations guarantee that the above equation holds for {\em all \/} values of $\phi_i$.   This means that a Taylor expansion of $H$ around the origin of field space will contain only a finite number of terms, and provide an exact solution to the truncated flow equations.  We  use this knowledge to write  $H$ as
\begin{equation}
\label{hgen}
H = \sum_{i,j,k\ldots} A_{ijk\ldots}\phi_1^i\phi_2^j\phi_3^k\ldots \, .
\end{equation} 
In (\ref{hgen}) the sum runs from $i+j+k+\ldots=0 $ to $i+j+k+\ldots =m$ so that the highest order terms in the expansion have {\it total} order $m$.

Given an expression for $H$, we can find the coefficients $A_{ijk\ldots}$ by taking appropriate partial derivatives.  For instance, for any number of fields $H$ contains a constant term $A_{000\ldots}$, which can be found by evaluating $H$ at $\phi_i =0$
\begin{equation}
\label{evalh}
H|_{\phi_i=0}=A_{000\ldots} = H_0.
\end{equation}
The value of $H_0$ was, of course determined by (\ref{hubblestart}).    

The higher order co-efficients may be best understood via a specific  example. For argument's sake,  consider the value of $A_{121}$ in a three field model:
\begin{equation}
2 A_{121} = \frac{\partial^4 H}{\partial \phi_1 \partial\phi_2 \partial\phi_2 \partial\phi_3}
\end{equation}
where the prefactor arises from the multiplicity in the partial derivatives.  
As we know
\begin{eqnarray}
^3\lambda_{1223}&=&\left(\frac{1}{4\pi G}\right)^3 \frac{1}{H^3}\left(\frac{\partial H}{\partial \phi_2}\right)\left(\frac{\partial H}{\partial \phi_3}\right)\frac{\partial^4 H}{\partial \phi_1\partial\phi_2\partial\phi_2\partial\phi_3}\\
&=& \left(\frac{1}{4 \pi G}\right)^{2}   \frac{\epsilon_2 \epsilon_3}{H}  \frac{\partial^4 H}{\partial \phi_1\partial\phi_2\partial\phi_2\partial\phi_3},
\end{eqnarray}
or when we evaluate at the origin, using (\ref{evalh}),
\begin{equation}
A_{121} = {A_{000}}\frac{\left({4 \pi G}\right)^{2}}{2} \frac{^3\lambda_{1223}}{\epsilon_2 \epsilon_3}.
\end{equation}

From these coefficients, we can solve for the explicit form of $\epsilon(\phi_i)$ by taking appropriate partial derivatives of $H(\phi_i)$:
\begin{equation}
\epsilon(\phi_i)=\left(\frac{1}{4 \pi G}\right) \frac{1}{H^2} \sum_{k=1}^M \left(\frac{\partial H}{\partial \phi_k}\right)^2.
\end{equation}
The full analytic form of the potential is  a complicated combination of these expansion coefficients (inverting (\ref{multihj}))
\begin{eqnarray}
V(\phi) &=& \frac{3}{8 \pi G} H^2(\phi)\left[{1-\frac{1}{3}\epsilon(\phi)}\right] \\
\label{manalytic}
&=& \frac{3}{8 \pi G} \left({\sum_{i,j,k\ldots z } A_{ijk\ldots z}\phi_1^i\phi_2^j\phi_3^k\ldots\phi_M^z} \right)^2\left[{ 1- \left(\frac{1}{12 \pi G}\right)  \sum_{k=1}^M \left(\frac{\partial_k \left({ \sum_{i,j,k\ldots z } A_{ijk\ldots z}\phi_1^i\phi_2^j\phi_3^k\ldots\phi_M^z}\right) }{\sum_{i,j,k\ldots z } A_{ijk\ldots z}\phi_1^i\phi_2^j\phi_3^k\ldots\phi_M^z }\right)^2 }\right]
\end{eqnarray}
Equation (\ref{manalytic}) provides an algorithm for finding an infinite number of {\it exact} inflationary solutions.   This is simply the multi field generalization of a process that has long been applied to the single field case  \cite{easther93,muslimov,salopekbond,lidsey91} -- there is of course no guarantee that the resulting solutions will correspond to well-motivated potentials.  Moreover, as noted previously we can transform the fields in any way that leaves the form of the kinetic terms unchanged.  In this case, the explicit form of $H(\phi)$ and $V(\phi)$ will change (since these functions are not invariant under these transformations) although the physics is obviously not affected.  Consequently, if one   wanted to implement a multi field reconstruction algorithm, one would need to take this ambiguity into account.   

\section{Cosmological Perturbations}

We now turn our attention to the power spectrum of scalar perturbations \cite{starobinskyper,sasakistewart} 
\begin{equation}
P_\mathcal{R}=\left(\frac{H}{2\pi}\right)^2\sum_{i=1}^M \left(\frac{\partial N}{\partial \phi_i}\right)^2 \, .
\end{equation}
Consistent with \cite{sasakistewart} we assume that $N$ is the number of {\it e}-foldings from a reference point {\it after} complete reheating.  A full discussion of isocurvature perturbations is beyond the scope of the current paper, and we do not include them here.   For a clear treatment discussion of the generation of isocurvature perturbations in multifield inflation see Bassett, Tsujiakawa and Wands' recent review \cite{Bassett:2005xm}.

This problem has been solved for a very general set of scalar fields which couple via their kinetic terms as well as through their potential  \cite{sasakistewart,lythriotto}.  However, since we have assumed canonical kinetic terms from the outset, we can find considerable simplifications to the general expressions -- in technical terms we are assuming that the metric on the scalar field space is flat, or that   $h_{ab}=\delta_{ab}$ in equation (45) of \cite{sasakistewart}}.  The spectral index $n_\mathcal{R}$ is given by
\begin{equation}
n_\mathcal{R}-1=\frac{d \ln P_\mathcal{R}}{d \ln k}
\end{equation}
or, again by \cite{sasakistewart, liddlemazumdar}, 
\begin{equation}
\label{index1}
n_\mathcal{R}-1=2\frac{\dot{H}}{H^2} - 2\frac{ \frac{dN}{d\phi_a} \left({8\pi G\frac{\dot{\phi}_a \dot{\phi}_b}{H^2} -  \frac{\partial^2 V}{\partial \phi_a \partial \phi_b} \frac{1}{8\pi G}\frac{1}{V} }\right) \frac{dN}{d\phi_b} } {\delta_{ij} \frac{dN}{d\phi_i}\frac{dN}{d\phi_j}}
\end{equation}
where we take advantage of summation convention.  To evaluate this for our specific models, we use two important substitutions:
\begin{eqnarray}
2\frac{\dot{H}}{H^2} & =& \frac{2}{H^2}\left({\frac{\ddot{a}}{a}-H^2}\right) \\
&=& -2{\epsilon}
\end{eqnarray}
and
\begin{equation}
\frac{\partial^2 V}{\partial \phi_b \partial \phi_a} = {3H^2}\left[{\epsilon_a \epsilon_b+\,^1\lambda_{ab}}\right]-H^2 \sum_{j=1}^M \left[{ \,^1\lambda_{bj} \,^1\lambda_{aj} + \,^2\lambda_{abj} }\right].
\end{equation}
Equation (\ref{index1}) becommes
\begin{equation}
\label{mspecindex}
n_\mathcal{R}-1  = -(2+4M)\epsilon +\frac{\epsilon}{1-\frac{1}{3}\epsilon} \sum_{a,b=1}^M \left({ 2\left[1+\frac{\,^1\lambda_{ab}}{\epsilon_a\epsilon_b}\right]-\frac{1}{3}\left(\sum_{j=1}^M\frac{{ \,^1\lambda_{bj} \,^1\lambda_{aj} + \,^2\lambda_{abj} }}{\epsilon_a\epsilon_b}\right)}\right).
\end{equation}
In the single field limit, $m=1$, we see that, to first order,
\begin{equation}
n_\mathcal{R}-1 = -4\epsilon + \,^2\lambda = \sigma
\end{equation}
which agrees with the approximation used in \cite{eastherkinney}.

The second observational parameter of interest is the ration of the amplitudes of the tensor to scalar perturbations, $r=T/S$.  For this we use the combined results: the power spectrum for gravitational waves can be expressed as \cite{sasakistewart} 
\begin{equation}
P_g=\left({\frac{H}{2\pi}}\right)^2
\end{equation}
and the spectral index for gravitational waves is given by \cite{sasakistewart,liddlelyth}
\begin{equation}
n_g=2\frac{\dot{H}}{H^2}=-2\epsilon,
\end{equation}
where the second equality translates the result into the slow roll parameters introduced in this paper.

The spectra obey a generalized consistency condition \cite{sasakistewart,polarskistarobinsky}, 
\begin{equation}
\frac{P_g}{P_{\mathcal R}}\leq|{n_g}| 
\end{equation}
which generalizes the consistency condition for inflation driven by a  single scalar field.   This ratio is proportional to the scalar:tensor ratio, $r$.

\section{Discussion}

In this paper, we have developed the Hubble slow roll formalism for  multi field models of inflation.  The properties of the multi field expansion are all easily understood as  generalizations of their single field counterparts \cite{liddleparsonsbarrow}.  In particular, we show that there is a hierarchy of differential (``flow'') equations that describe the evolution of the slow roll parameters. If only a finite number of terms in the hierarchy are non-zero at a given instant then the flow equations ensure that this truncation is preserved as the universe evolves.  

In principle, one could generate a large set of multi field models and attempt to constrain the slow roll parameters with observational data,  as was done for the single field case via Monte Carlo reconstruction  \cite{eastherkinney}.   We  briefly examined this possibility, but it appears that the larger number of parameters  at each order in the multi field slow roll expansion makes it unlikely we can put meaningful constraints on the multi field parameter space for any realistic dataset. Moreover, while we have assumed that our fields have minimal kinetic terms, we can still rotate the fields into one another, preserving the form of the kinetic terms but modifying the functional form of the potential, and any reconstruction algorithm  would have to take this ambiguity into account.

While Monte Carlo reconstruction may not be practical in the multi field setting, Liddle, Parsons and Barrow's systematic treatment of single field inflation in terns of the Hubble slow roll expansion has found a large range of applications in thorough studies of the dynamical system that underlies these models. We consequently expect that this multi field generalization is a theoretical tool  with a  number of useful applications. We have shown there that there is a strong correspondence between the two systems and that many features of the single field case generalize naturally to their multi field counterpart.   In particular,  the truncated multi field flow equations can be solved exactly, yielding an expression for the potential analogous to that  obtained from  Liddle's exact solution to the single field hierarchy \cite{liddle03}.  If one wished to do so, this solution to the flow equations could be used  to generate any number of exact multi field inflationary solutions.   Moreover, we can write down expressions for the perturbation spectrum using the multi field slow roll parameters, and in future work we plan to use this formalism to examine the perturbation spectrum produced when the inflaton does not follow  a single, well-defined path in the multi dimensional inflationary potential.

\section{Aknowlegements}

RE and JTG are supported in part by the United States Department of Energy, grants DE-FG02-92ER-40704 and DE-FG02-91ER40688, TASK A respectively. We thank Will Kinney for a number of useful conversations about this topic.


\begin{thebibliography}{0}
\bibitem{guth}
  A.~H.~Guth,
  ``The Inflationary Universe: A Possible Solution To The Horizon And Flatness  Problems,''
  Phys.\ Rev.\ D {\bf 23}, 347 (1981).

\bibitem{linde82}
  A.~D.~Linde,
  ``A New Inflationary Universe Scenario: A Possible Solution Of The Horizon,
  Flatness, Homogeneity, Isotropy And Primordial Monopole Problems,''
  Phys.\ Lett.\ B {\bf 108}, 389 (1982).

\bibitem{steinhardtturner}
  P.~J.~Steinhardt and M.~S.~Turner,
  ``A Prescription For Successful New Inflation,''
  Phys.\ Rev.\ D {\bf 29}, 2162 (1984).

\bibitem{liddleparsonsbarrow}
  A.~R.~Liddle, P.~Parsons and J.~D.~Barrow,
  ``Formalizing the slow roll approximation in inflation,''
  Phys.\ Rev.\ D {\bf 50}, 7222 (1994)
  [arXiv:astro-ph/9408015].

\bibitem{starob1978}
   Starobinskii, A.~A. 
   ``On a nonsingular isotropic cosmological model''
   Sov.\ Astron.\ Lett. {\bf 4} 82 (1978)

\bibitem{kinney02}
  W.~H.~Kinney,
  ``Inflation: Flow, fixed points and observables to arbitrary order in  slow
  roll,''
  Phys.\ Rev.\ D {\bf 66}, 083508 (2002)
  [arXiv:astro-ph/0206032].

\bibitem{eastherkinney}
  R.~Easther and W.~H.~Kinney,
  ``Monte Carlo reconstruction of the inflationary potential,''
  Phys.\ Rev.\ D {\bf 67}, 043511 (2003)
  [arXiv:astro-ph/0210345].

\bibitem{wmap}
  C.~L.~Bennett {\it et al.},  
  ``First Year Wilkinson Microwave Anisotropy Probe (WMAP) Observations:
  Preliminary Maps and Basic Results,''
  Astrophys.\ J.\ Suppl.\  {\bf 148}, 1 (2003)
  [arXiv:astro-ph/0302207].

\bibitem{peiris}
  H.~V.~Peiris {\it et al.},
  ``First year Wilkinson Microwave Anisotropy Probe (WMAP) observations:
  Implications for inflation,''
  Astrophys.\ J.\ Suppl.\  {\bf 148}, 213 (2003)
  [arXiv:astro-ph/0302225].

\bibitem{fengbo}
  C.~Y.~Chen, B.~Feng, X.~L.~Wang and Z.~Y.~Yang,
  ``Reconstructing large running-index inflaton potentials,''
  Class.\ Quant.\ Grav.\  {\bf 21}, 3223 (2004)
  [arXiv:astro-ph/0404419].

\bibitem{Efstathiou:2005tq}
  G.~Efstathiou and K.~J.~Mack,
  arXiv:astro-ph/0503360.

\bibitem{liddle03}
  A.~R.~Liddle,
  ``On the inflationary flow equations,''
  Phys.\ Rev.\ D {\bf 68}, 103504 (2003)
  [arXiv:astro-ph/0307286].

\bibitem{ramirezliddle}
  E.~Ramirez and A.~R.~Liddle,
  ``Stochastic approaches to inflation model building,''
  arXiv:astro-ph/0502361.

\bibitem{copelandliddle}
  E.~J.~Copeland, A.~R.~Liddle, D.~H.~Lyth, E.~D.~Stewart and D.~Wands,
  ``False vacuum inflation with Einstein gravity,''
  Phys.\ Rev.\ D {\bf 49}, 6410 (1994)
  [arXiv:astro-ph/9401011].

\bibitem{linde94}
  A.~D.~Linde,
  ``Hybrid inflation,''
  Phys.\ Rev.\ D {\bf 49}, 748 (1994)
  [arXiv:astro-ph/9307002].

\bibitem{barnard}      
  M.~Barnard and A.~Albrecht,
  ``On open inflation, the string theory landscape and the low CMB
  quadrupole,''
  arXiv:hep-th/0409082.

\bibitem{burgess}
  C.~P.~Burgess, R.~Easther, A.~Mazumdar, D.~F.~Mota and T.~Multamaki,
  ``Multiple inflation, cosmic string networks and the string landscape,''
  arXiv:hep-th/0501125.

\bibitem{easther04}
  R.~Easther,
  ``Folded inflation, primordial tensors, and the running of the scalar
  spectral index,''
  [arXiv:hep-th/0407042].

\bibitem{kklmmt}
  S.~Kachru, R.~Kallosh, A.~Linde, J.~Maldacena, L.~McAllister and S.~P.~Trivedi,
  ``Towards inflation in string theory,''
  JCAP {\bf 0310}, 013 (2003)
  [arXiv:hep-th/0308055].

\bibitem{salopekbond}
  D.~S.~Salopek and J.~R.~Bond,
  ``Nonlinear Evolution Of Long Wavelength Metric Fluctuations In Inflationary
  Models,''
  Phys.\ Rev.\ D {\bf 42}, 3936 (1990).

\bibitem{copelandkolb}
  E.~J.~Copeland, E.~W.~Kolb, A.~R.~Liddle and J.~E.~Lidsey,
  ``Reconstructing the inflation potential, in principle and in practice,''
  Phys.\ Rev.\ D {\bf 48}, 2529 (1993)
  [arXiv:hep-ph/9303288].

\bibitem{liddlelyth}
  A.~R.~Liddle and D.~H.~Lyth,
  ``COBE, gravitational waves, inflation and extended inflation,''
  Phys.\ Lett.\ B {\bf 291}, 391 (1992)
  [arXiv:astro-ph/9208007].

\bibitem{nibbelink1}
  S.~Groot Nibbelink and B.~J.~W.~van Tent,
  ``Density perturbations arising from multiple field slow-roll inflation,''
  [arXiv:hep-ph/0011325].

\bibitem{nibbelink2}
  S.~Groot Nibbelink and B.~J.~W.~van Tent,
  ``Scalar perturbations during multiple field slow-roll inflation,''
  Class.\ Quant.\ Grav.\  {\bf 19}, 613 (2002)
  [arXiv:hep-ph/0107272].

\bibitem{easther93}
  R.~Easther,
  ``Exact superstring motivated cosmological models,''
  Class.\ Quant.\ Grav.\  {\bf 10}, 2203 (1993)
  [arXiv:gr-qc/9308010].

\bibitem{lidsey91}
  J.~E.~Lidsey,
  ``The Scalar field as dynamical variable in inflation,''
  Phys.\ Lett.\ B {\bf 273}, 42 (1991).

\bibitem{muslimov}
  A.~G.~Muslimov,
  ``On The Scalar Field Dynamics In A Spatially Flat Friedman Universe,''
  Class.\ Quant.\ Grav.\  {\bf 7}, 231 (1990).

\bibitem{starobinskyper}
  A.~A.~Starobinsky,
  ``Multicomponent De Sitter (Inflationary) Stages And The Generation Of
  Perturbations,''
  JETP Lett.\  {\bf 42}, 152 (1985)
  [Pisma Zh.\ Eksp.\ Teor.\ Fiz.\  {\bf 42}, 124 (1985)].

\bibitem{Bassett:2005xm}
  B.~A.~Bassett, S.~Tsujikawa and D.~Wands,
  arXiv:astro-ph/0507632.

\bibitem{sasakistewart}
  M.~Sasaki and E.~D.~Stewart,
  ``A General analytic formula for the spectral index of the density
  perturbations produced during inflation,''
  Prog.\ Theor.\ Phys.\  {\bf 95}, 71 (1996)
  [arXiv:astro-ph/9507001].

\bibitem{lythriotto}
  D.~H.~Lyth and A.~Riotto,
  ``Particle physics models of inflation and the cosmological density
  perturbation,''
  Phys.\ Rept.\  {\bf 314}, 1 (1999)
  [arXiv:hep-ph/9807278].

\bibitem{liddlemazumdar}
  A.~R.~Liddle, A.~Mazumdar and F.~E.~Schunck,
  ``Assisted inflation,''
  Phys.\ Rev.\ D {\bf 58}, 061301 (1998)
  [arXiv:astro-ph/9804177].


\bibitem{polarskistarobinsky}
  D.~Polarski and A.~A.~Starobinsky,
  ``Structure of primordial gravitational waves spectrum in a double
  inflationary model,''
  Phys.\ Lett.\ B {\bf 356}, 196 (1995)
  [arXiv:astro-ph/9505125].

\bibitem{liddlebook}
  A.~R.~Liddle and D.~H.~Lyth,
  ``Cosmological inflation and large-scale structure,''

\bibitem{infreconstruct}
  J.~E.~Lidsey, A.~R.~Liddle, E.~W.~Kolb, E.~J.~Copeland, T.~Barreiro and M.~Abney,
  ``Reconstructing the inflaton potential: An overview,''
  Rev.\ Mod.\ Phys.\  {\bf 69}, 373 (1997)
  [arXiv:astro-ph/9508078].

\end{thebibliography}
\end{document}